# Intelligent Low-level RF System by Non-destructive Beam Monitoring Device for Cyclotrons

M.  S. Sharifi Asadi Malafeh [1], M. Ghergherehchi[1, 2a]

, H. Afarideh[1], J.S. Chai[2]

*[1]Department of Energy Engineering & Physics, Amirkabir University of Technology, Tehran, Iran*
*[2]College of Electronic and Electrical Engineering, Sungkyunkwan University, Suwon, Korea*

**Abstract** The project of a10MeV PET cyclotron accelerator for medical diagnosis and treatment was started at Amirkabir University of Technology in 2012. The low-level RF system of cyclotron accelerator is designed to stabilize acceleration voltage and control the resonance frequency of the cavity. In this work Intelligent Low Level Radio Frequency Circuit or ILLRF suitable for Most of the AVF cyclotron accelerators was designed by the beam monitoring device and narrow band tunable band-pass filter. In this design, for the RF phase detection does not need to signal processing by microcontroller.

---
ᵃ mitragh@skku.edu

## 1. Introduction

The tasks of ILLRF circuit are signal generation, phase and frequency adjustment, stabilization and protection of the RF set from the reflected power due to changing the cavity resonance frequency by thermal cavity deformation [1]. Therefore, in order to establish suitable electric field in cavities, a noiseless RF signal must be generated, amplified and applied to the cavity. Noise and signal harmonics are inseparable parts of all high-frequency electronic circuits; therefore, an appropriate filter must be used to eliminate harmonics and noise [2]. Regarding the sensitivity of the system to the harmonics of signal and noise, the narrow band tunable band-pass filter should be used. In cyclotron, variation of cavity resonant frequency causes the reflected power. Accordingly, for tune the resonant frequency of cavity and to avoid the risk, the reflected power detector (directional coupler) used.

When the variation of cavity resonant frequency is negligible, the best way to detect resonant frequency variations is measure the phase difference of the forward signal and the sampled signal of the Dees with using the RF phase detector. In regard to the relation between beam rotational phase and RF signal phase, the lack of coherence between them will cause a loss of beam resolution. In this work we proposed a method to solve this problem, which use beam monitoring device to derive the phase, position and current information of the beam. Also In this work instead of using signal processing by microcontroller, we used only simple microcontroller for the RF phase detection.

In designing ILLRF, if RF signal phase detected by signal processing in the processor; the main problem will be to choose the appropriate processor. However, in this design, the signal processing is eliminated and RF phase is detected by a special IC which is the most important part for the ILLRF amplitude and phase stability, protection and cavity resonant frequency tuning.

Using narrow band-pass filter to reduce the effects of noise and harmonics causes to reducing the overall bandwidth of system. Therefore, the LLRF with fixed narrow band-pass filter is designed for only one specific accelerator. But in this design, by using the tunable band-pass filter with an ultra-narrow-band-pass, the general bandwidth of the system does not decrease. The ILLRF circuit because of using appropriate sub-blocks and the prediction of all the probable conditions demonstrates a high level of intelligence. The general Schematic of RF section is shown in Fig.1.ILLRF circuit is start-up block of RF section and consists of the Signal Generation and Adjustment Block and with help of Amplitude, Frequency and Phase Feedback Loops and Beam Monitoring Device, Protect the system and adjustment the RF signal. ILLRF circuit consists of the following main Blocks:

1- Signal Generation and Adjustment Block

2- Amplitude, Reflected power and Phase Feedback Loops

3- Protection Block

4- Automatic gain control Block for amplitude stability

5- Beam Monitoring Device for Beam Phase, Current, and Position Detection

The ILLRF with feedback's, output and Beam monitoring device, is shown in Fig.2.

Taking into account the RLC equivalent circuit of the cavity, if the input signal frequency is different from the cavity resonant frequency, we will have the reflected power that can be dangerous at the maximum power. However, the RLC equivalent circuit of the cavity means input impedance that is seen from the high power transmission line and consists of coupling capacitor, cavity and tuning capacitor.

## 2. Design of intelligent low-level RF system

The reflected power will cause to decrease signal amplitude in the Dee's. Consequently, to analyze the cavity resonant frequency, in addition to measuring the reflected power and phase Variations (the phase difference of Dee's pick-up voltage and forward signal), we should increase the assurance factor by measuring the signal amplitude of Dee's.

If there is an imbalance between phase of the cavity interior electric field and phase of the particle motion, we will not have exterior beam.

As result, to use ILLRF next to the pulsed ion source, the utilization of beam monitoring device is vitally important.

According to flowchart in Fig.3, by feedback Loops, the microcontroller controls the ILLRF. In the initial configuration, before the startup, the designed working frequency of cyclotron will be sent to the microcontroller through the serial interface.

## 2.1 Signal Generation

The input clock frequency for DDS will be established, by temperature controlled crystal oscillator (TCXO) crystal oscillator and the output frequency is a fraction of the input clock frequency [3]. The path of the Generation and the adjustment of RF signal and the protection of the system can be observed in the block diagram of Fig.4.

The Microcontroller sends the commands for changing the phase and frequency of DDS by SPI. Analog signal in DDS will be produced digitally. Therefore, the output consists of the fundamental signal and its harmonics. ILLRF has two filters to eliminate noise and harmonics. First, low-pass filter after the DDS output [4] and second narrow-band tunable band-pass filter after preamplifier to remove amplified Residual noise and harmonics. The amplified signal amplitude stability in Dee's related to LLRF and power amplifier amplitude stability, so according to the block diagram of Fig.4, the Dee's pick-up signal amplitude by ADC, Recorded in microprocessor and local microprocessor report it to the main processor that shown in working flowchart of Fig.3. Simultaneously to control the output level and amplitude stability after RF to DC Demodulator goes to AGC of Preamplifier.

## 2.2 Startup Process

In the startup process, according to multipactoring effect and to prevent mismatching, sparks and instability condition during the startup process, first the microcontroller will set the output signal amplitude at the lowest level with 10 percent duty cycle and by measuring the phase difference of Dee's pick-up voltage and amplified signal, the tuning capacitor adjusted. Then slowly with an appropriate slope increased duty cycle to 100 percent and power to maximum and then the Dee's signal amplitude, the reflected power and the phase difference will also be analyzed and the readjustment of resonant frequency will be made.

## 2.3 Protection and Readjustment Process

After the complete startup, the tasks of ILLRF will be the non-stop analysis of the accuracy of the task, protection, the correction of initial settings and transmitting the data of beam and signal to the main processor of the cyclotron. In protection part when the reflected power is more than 50 Watts and less than 300 watts to prevent the devastating effects of the reflected power, the amplitude of amplified signal was decreases and then after correction the resonant frequency of the cavity, again was increases. If the cavity resonant frequency is not corrected then the Critically level of phase and amplitude and reflected power are diagnosed. If Variations in phase is more than 11.5 degree or Increase the reflected power more than 300 w, the signal is deflected to the dummy load by high power switch.

## 2.4 Amplitude Stability

Amplitude stability of amplified signal related to ILLRF amplitude stability and Power-Amplifier amplitude stability, so the two errors will decrease the sustainability, therefore the ILLRF compensates the amplitude error of power-amplifier by Dee amplitude feedback and AD8367 automatic gain control.

*2.5 Variations and correction of Resonance Frequency*

For the Frequency difference detection between the cavity resonant frequency and the RF signal frequency, sampling procedure of the forward and reflected power will be established by directional-coupler and the Dee's signal is sampled by an antenna and these signals in both digital and analog paths will be analyzed. While in recently reported work by Pengzhan Li, et al, reflected power and Dee amplitude signals are used only for protection [5]. Main path of resonance frequency correction is the analogue path and digital correction is for safety and stability. In analogue path phase difference of Dee's pick-up signal by the sampling antenna and amplified signal by directional-coupler are detected by the phase detector AD8302 [6] and directly tune the tuner capacitor by servo motor. On the digital path, the Dee's amplitude and reflected the Maximum range of analogue correction is 2 degree so if microcontroller detects the reflected power more than 50 W equal to 2 degree (According to the diagram in Fig.5) will digitally adjust the cavity resonant frequency by tuning capacitor. On the protection path, the amplitude of the reflected power will be compared with the level of the reference reflected power uninterruptedly.

Taking into account the RLC equivalent circuit of the cavity and RF signal frequency, the reflected power levels by Eq.1 are easily attainable.

$$P_{ref} = P_{fwd} - P_{loss} = P_{fwd} - P_{fwd} \times \left| \frac{R}{R + \left( LW - \frac{1}{CW} \right) j} \right| \quad (1)$$

By using reflected power just could be realized to the amount of the cavity resonant frequency variation but according to the diagram in Fig.5 by using the phase difference between the cavity pick-up signal and forward power, resonant frequency variation is determined by the magnitude and sign. Due to Specification of the AD8302 phase detector [6], as shown in Fig.6, best response is in ±90 degrees.

Phase Difference of forward power and RF cavity signal in resonance frequency is the reference phase. So by phase shifter, the reference phase is transferred to 90 degree.

Except for the reflected Power and the phase difference the third method to resonance frequency variation detection is the analysis of signal amplitude inside the Dee's, which is the last option for the protection of the RF set.

*2.6 Working Flowchart*

Depending on the variation rate of the resonant frequency, the error signal will be divided into 4 levels:

**First Level:** zero to 50w of the reflected power equal to proximately 2 degree phase difference According to the diagram in Fig.5. It is equal to ±1MHz difference from the working frequency. After detection of resonance frequency variation, analogue path tune the cavity resonance frequency by tuning capacitor. For the elimination of errors in the first level, there is no need to change the transmitted power because the reflected power is very little and it is not considered to be of any danger.

**Second level:** 50 to 300 w of the reflected power

At first through the proper slope, the level of the signal amplitude is decreased and then, the cavity resonant frequency is adjusted Digitally, but if the digital path is unable to detect the change in the resonant frequency or to make the necessary adjustment; to avoid the probable danger, it minimizes the amplitude of signal with a steep slope by the analog path.

**Third level:** More than 300 watts of the reflected power equal to 11.5 degree phase difference. This level is 2 percent of forward power and equal to ±2.25MHz difference from the operating frequency, for the emergency protection of the power amplifier. Therefore, the level of signal amplitude with a steep slope has independently reached the minimum and the error signal for restarting is transmitted to the microcontroller.

The amplitude inside the Dee's has been used in both analog and digital paths. In analog path, the Dee's signal amplitude will be compared with the reference Dee voltage for the protection. In the digital path, for the detection of the frequency variation rate, a factor of the signal amplitude is digitalized by the microcontroller and will tune the cavity resonant frequency by tuning capacitor. According to the flowchart in Fig.3, Such as reflected power detection, also in Dee pick-up voltage detection, two threshold levels considered that equals to 2 and 11.5 degrees.

## 3. The principle of the non-destructive beam monitoring device

The task of this device is to derive time, position and current information of the beam simultaneously. The best time for accelerating a particle is the 45 degree phase of RF signal. Therefore, the difference between the beam turning phase and the RF signal phase indicates the proper or the improper operation of the cyclotron. Consequently, if we do not have the output beam, the last beam current and phase information's are more important in maintenance procedure. In order to produce the isochronous fields, this structure has an advantage that it can be easily installed in the cyclotron. As a result, for the initial setup and complete control over the operation of the system, we need to detect the beam current. Beam current detection with the help of FARADAY CUP causes the blockage of the exit Beam, but this detector by locating the beam probe in the valleys as shown in Fig.7, during the whole operation of accelerator, detects the beam without causing any blockage [7].

The charged particle generates a magnetic field while it is moving (as shown in Fig.8). The amount of this field according to Eq.2 is relative to the charge of the particle, the distance from the path and the magnetic permittivity coefficient [8].

$$\vec{B} = \frac{\mu_0}{4\pi} \frac{q\vec{V} \times r\widehat{r_P}}{r_{P_P}^2} \qquad (2)$$

Where r is the distance in meters, V is the speed in meters per second and q is the charge in clone. The number of particles in each bunch of cyclotron accelerators is about $1e^6 \sim 1e^8$ particles. For example, we assume the 5MeV proton. To calculate its magnetic field, according to Eq.3, first the speed of each particle should be calculated:

$$KE = \frac{1}{2} \frac{m}{\sqrt{1 - \left(\frac{V^2}{c^2}\right)}} V^2 = \frac{E}{6.241509 \times 10^{18}} \text{ Joule}$$

$$V = \frac{(1.41 \times \sqrt{(KE \times \sqrt{(KE^2 + c^4 \times m^2)} - KE^2))}}{c \times m} = 3.0864 \times 10^7 \, {}^m\!/_S \qquad (3)$$

Where m is the mass of particle in gram (m= $1.673 \times 10^{-27}$g), C is the speed of light, E is the energy of the particle, M is the particle relativistic mass and KE is the kinetic energy. According to Eq.3, the speed of 5MeV proton will be equal to $3.0864 \times 10^7$m/S. And by Eq.2 we get the magnetic field.

Considering the equation I=-N $\Delta\varphi/\Delta$tt, the amount of the induced current in the beam monitoring probe will be assessed, also with the help of Eq.4, magnetic flux in each spot will be calculated and by working out the flux in two spots, $\Delta\varphi$ will be attained. When the speed of the particle is clarified, the necessary time for moving the particle from the first point to the second one will become equal to $\Delta$t and consequently for this purpose, having a coil with the area of 4 $cm^2$, the maximum amplitude of the induced current will become equal to 28 µA.

$$\varphi = B.A.cos(\theta) \qquad (4)$$

where B is the magnetic field and A is the area of the coil. In calculations, contrary to simulation, for simplification of the beam bunch, a spot beam will be considered. Therefore, the amplitude of the induced current will become a little more. Using simulation in CST PARTICLE STUDIO, the induced current in the coils of the beam monitoring probe, with different cross section, as shown in Fig.9 and with equal areas was studied. By investigating of literature, only one work on nondestructive phase detection has been reported [7]. Although this work relatively has good contribution unfortunately it is not accurate. According to the graphs shown in Fig.10 obtain by particle studio of CST software [9], Contrary to the PHASE PROBE cross section reported by Satoru Hojo, et al [7], the optimal cross section form for coils is the circular cross section because its output signal has the highest induced current and the fastest zero crossing.

The graphs in Fig.11 show the induced current in single-looped coils with different distances compared to the path of the particle whose maximum induced current is about 18μA. As the coils move some distance from the beam's path, the induced current decreases and consequently this situation is used for the detection of the beam motional radius. Zero crossing time will be utilized as the reference for timing. To detect the amount of the beam's current at any moment, the amount of the induced current of coils will be added up after being rectified. By the calibration, the total induced current will be attributed to the amount of the beam's current.

## 4. The outline of some main sub-blocks of ILLRF

### 4.1 Variations and correction of Resonance Frequency

Because of using appropriate sub-blocks, the level of processing in the microcontroller is decreased, so there is no need to employ very high speed microcontrollers; only the microcontroller of Atmel AVR AT-mega family is enough, taking into account the necessary measures for the reduction of noise. While in recently reported work by Pengzhan Li, et al, using DSP is over-designed [5]. To send necessary data and non-automated control by the operator, the microcontroller is connected to the main processor by serial interface.

### 4.2 Direct Digital Synthesizer

DDS by TCXO crystal oscillator has the responsibility of signal generation and adjustment of signal phase and frequency. In this design instead of VCOs and I-Q modulators AD9859 Synthesizer used so the phase and frequency easily adjusted by microcontroller [3].

### 4.3 The VGA Preamplifier

For stable signal amplitude and the RF set protection, preamplifier with automatic gain control , variable gain amplifier AD8367 is utilized (as shown in Fig.12). [10].

### 4.4 Filters

DDS generates the analog signal by digital techniques. Therefore, the harmonics of generated signal will also appear. By using a suitable low-pass filter the DDS output signal is quite clean and non-harmonic [2]. To decrease the noise and the left-over harmonics and for more control and protection, the ultra-narrow-band tunable band-pass filter will be used after preamplifier as shown in Fig.13.

### 4.5 Demodulator

To compare the power and amplitude of the signal with the reference power and amplitude, employing the signal amplitude is adequate. Therefore to reduce the unwanted errors, first the level of the sampled signal with a fixed factor is converted to a usable one and then with the help of chips AD8361 [11] and AD736 [12] RF signal is converted to a DC voltage level.

### 4.6 The Beam Monitoring Device Description

To detect the turning radius, the beam monitoring probe that consists of 88 coils is used. Therefore, in Fig.14, M is equal to 88.The induced current of each coil in a high sensitive current to voltage amplifier converts into a voltage signal with appropriate amplitude [13]. Then the signal is guided into two paths to acquire time-phase data and the current -position data. For the beam Characteristics calculation, an auxiliary processor is used to decrease processing level of the main microcontroller of ILLRF. On the obtained time data path, the zero crossing time is the reference of timing. Consequently, the zero crossing time of Biopsied RF signal used to startup the Time-to-Amplitude Converter or TAC. Then, the induced current zero crossing time will stop the TAC.

To reduce the noise effects, between the induced current zero crossing detector and the induced current peak detector, timing is synchronized. The output amplitude of TAC is converted to digital data and sent to the auxiliary microcontroller.

Table 1. The result of RF test and some characteristics of this cyclotron.

| Parameter | value |
|---|---|
| Operating RF frequency | 70MHz |
| RF output power | 15 KW |
| Magnetic field between upper and lower magnet pole | 0.26 up to 1.83T |
| RF frequency tolerance @ -40 to +85°C | ±1.75 KHz |
| RF frequency tolerance @ +25°C | ±350 Hz |
| Dee voltage stability | ±7×10⁻⁴ |
| Max VSWRMaximum B field | <1.3179 |
| Phase error | <±0.63° |
| Minimum detectable Resonance frequency variation | <200 KHz |
| Operating temperature | -40° to +85° C |
| Typical temperature | 25° C |
| Second-Order Harmonic Distortion | -58dBc |

To acquire current data, the rectified amplitude of the coils is collected by SUM unite and after digitalization is sent to the auxiliary microcontroller. On the path of extracting position-radius data, the coil with the highest amplitude of the induced current represents the motion radius of the beam. Therefore, the maximum amplitude is detected by a comparator and the number of coil with maximum amplitude is sent to the auxiliary microcontroller as the indicator of the motion radius. For this purpose, the 88 wire Parallel BUS is required and proportional to it, the 100-pin microcontroller is needed, but based on Fig.15, by using Multichannel Priority Encoder Comparator and Multichannel Programmable Linear Gate, the connecting path of the coils to the auxiliary microcontroller is reduced to a 17 wires BUS by SN74HC148 8-Line To 3-Line Priority Encoders [14].

5. Conclusion

The proper circuit was designed to minimize the level of noise and harmonics, by tunable narrow band-pass filter for a wide range of RF accelerators which has no limitation of lessening the bandwidth of the output as exist in telecommunication systems when based on synthesizers. Protecting the bandwidth is a vital matter when ILLRF going to be used by wide range of RF accelerators and it is absolutely preserved in this work.

The open loop RF generation and adjustment have been tested and the result of RF test and some characteristics of this cyclotron have been demonstrated in Table.1.

Figure 1

The general Schematic of RF section

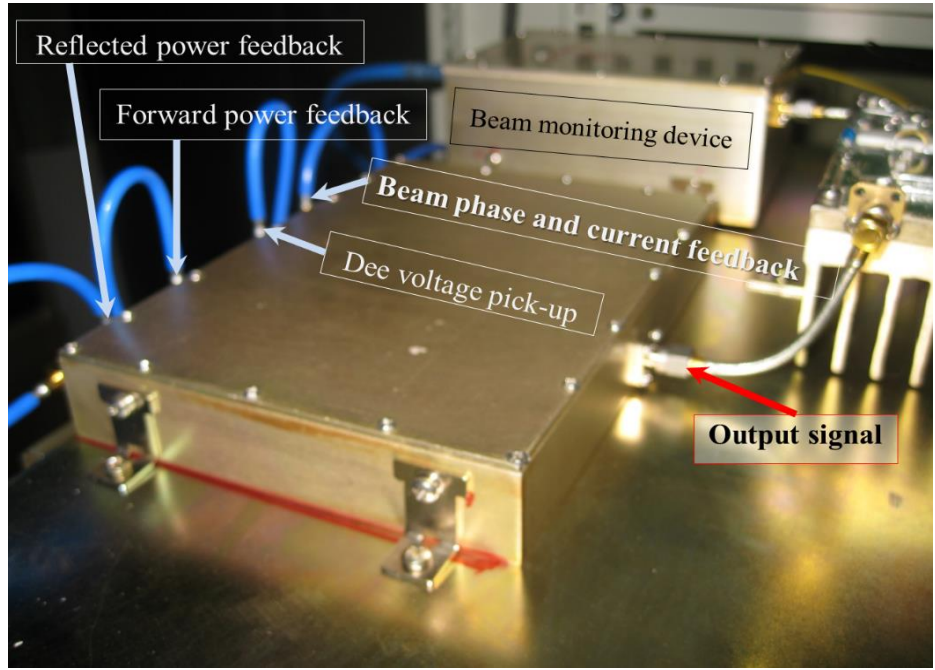

**Figure 2**

The ILLRF with feedback's, output and Beam monitoring device

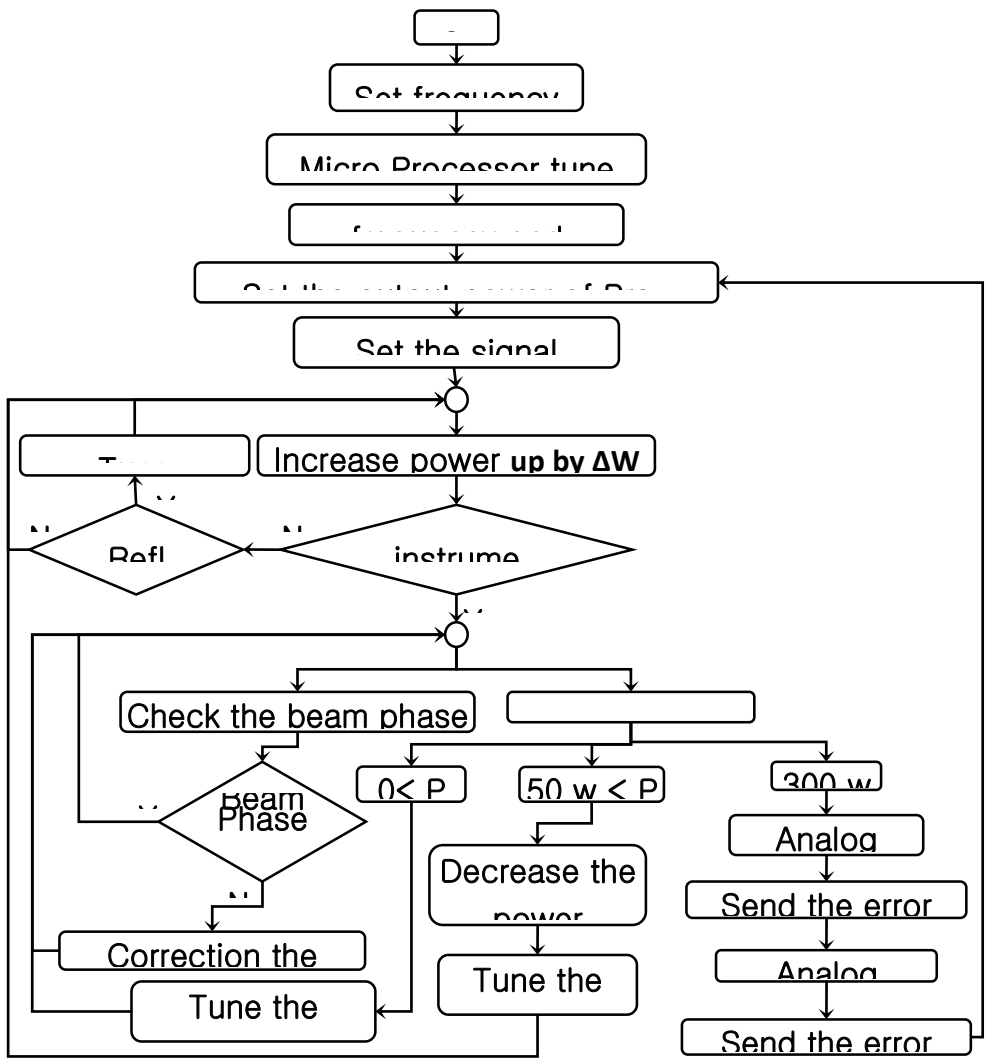

Figure 3

ILLRF operation flowchart

**Figure 4**

ILLRF circuit block diagram and its connection to other parts of the RF set

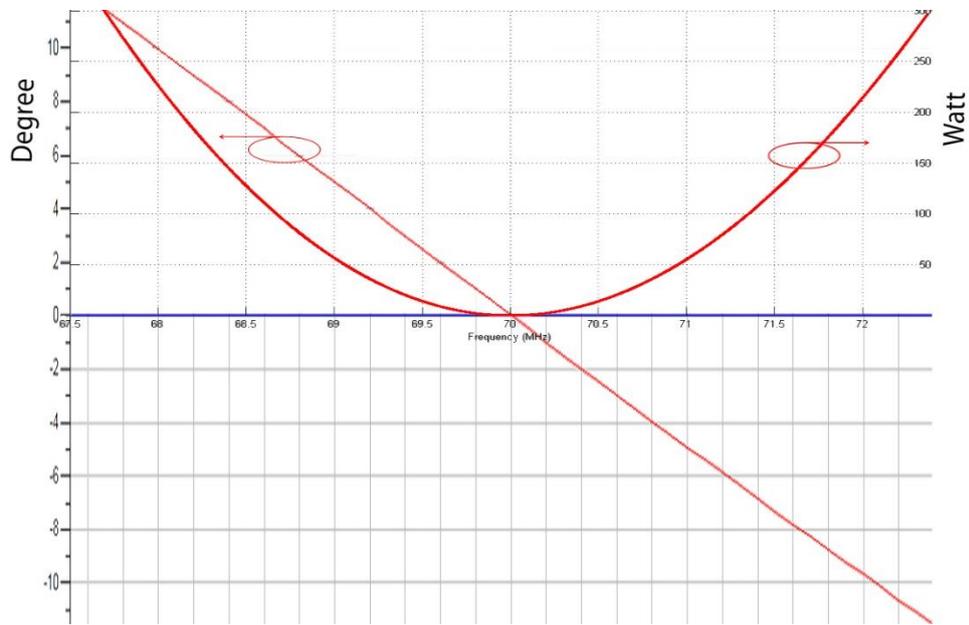

**Figure 5**

The relationship between phase difference and reflected power in various frequencies

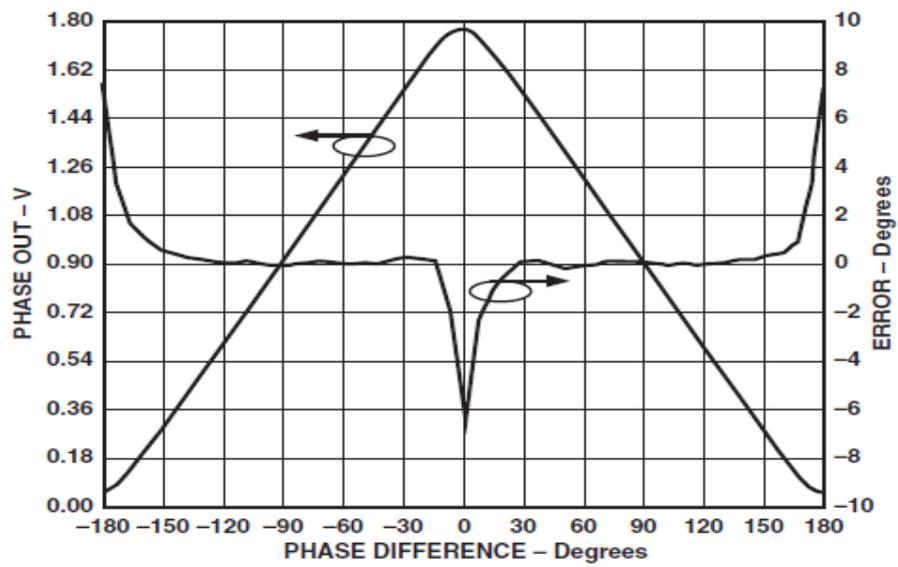

Figure 6

VPHS Output and Nonlinearity vs. Input Phase Difference, Input Levels −30 dBm, Frequency 100 MHz

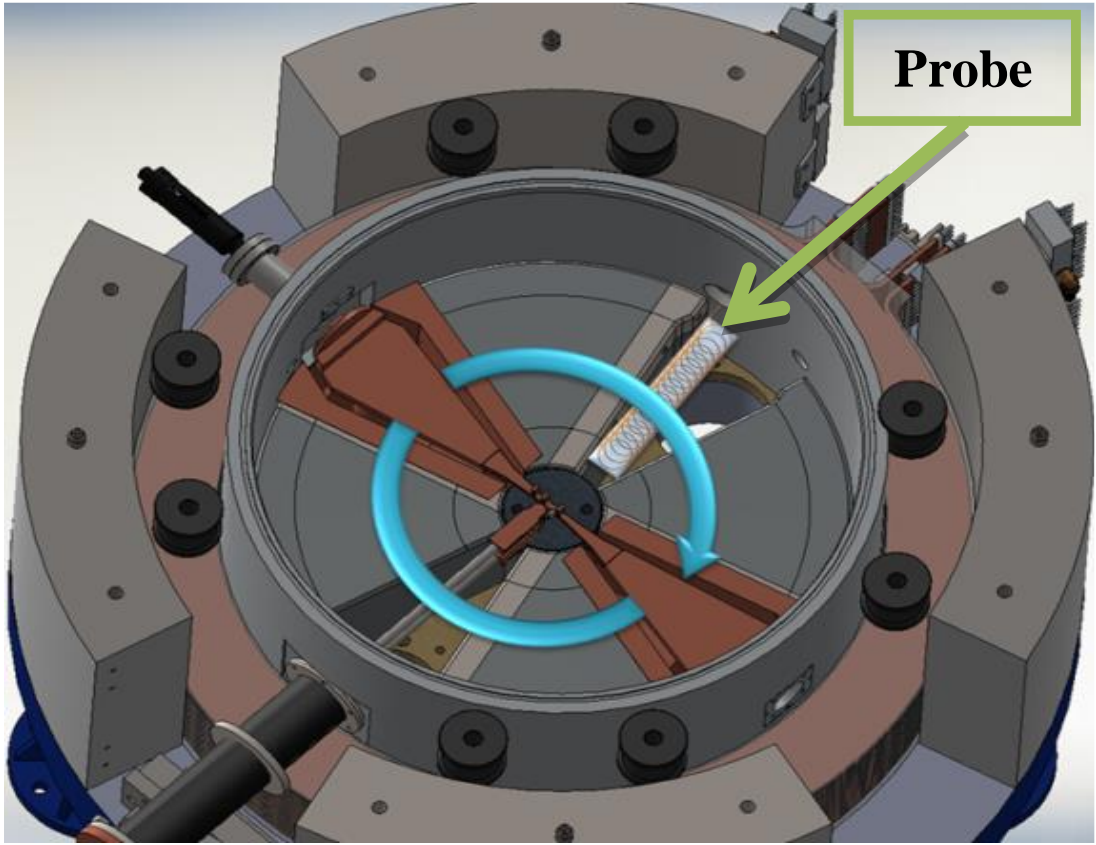

**Probe**

Beam monitoring probe locating in the space valleys

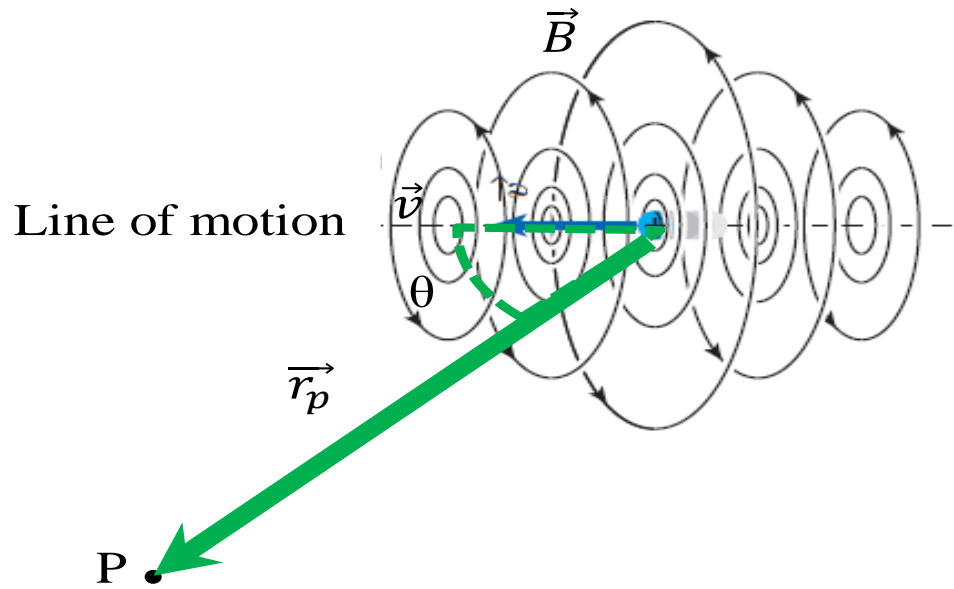

Line of motion

Figure 8

Magnetic field caused by the motion of charged particles

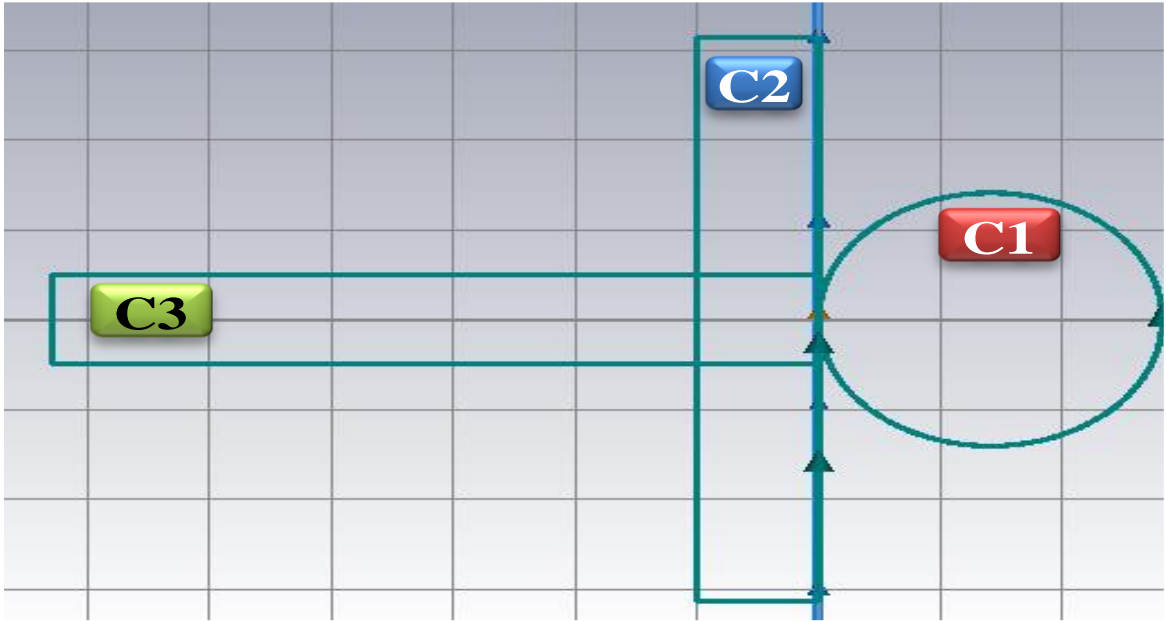

Coils with different cross sections

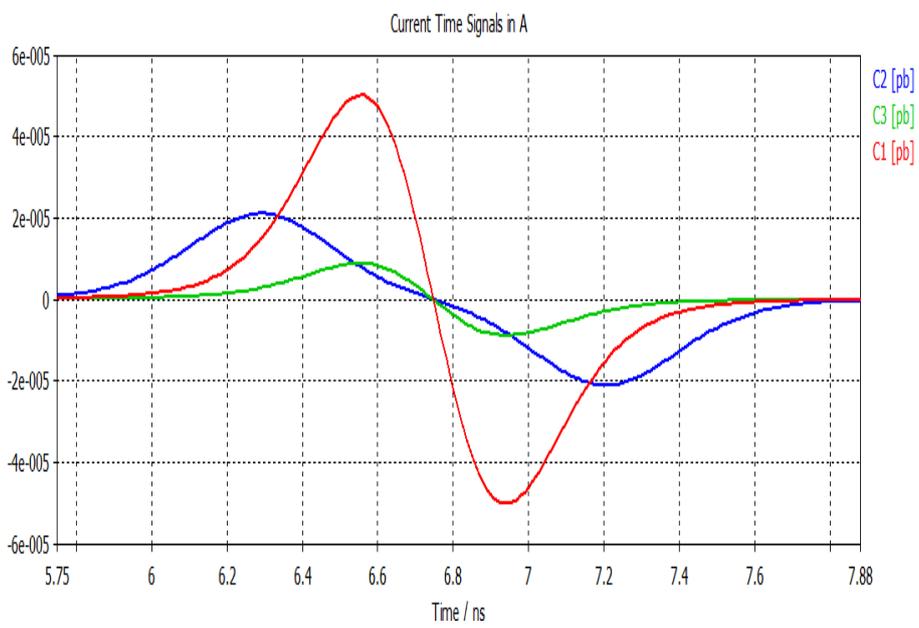

Induced current in the coils of beam monitoring probe with different cross-sectional areas

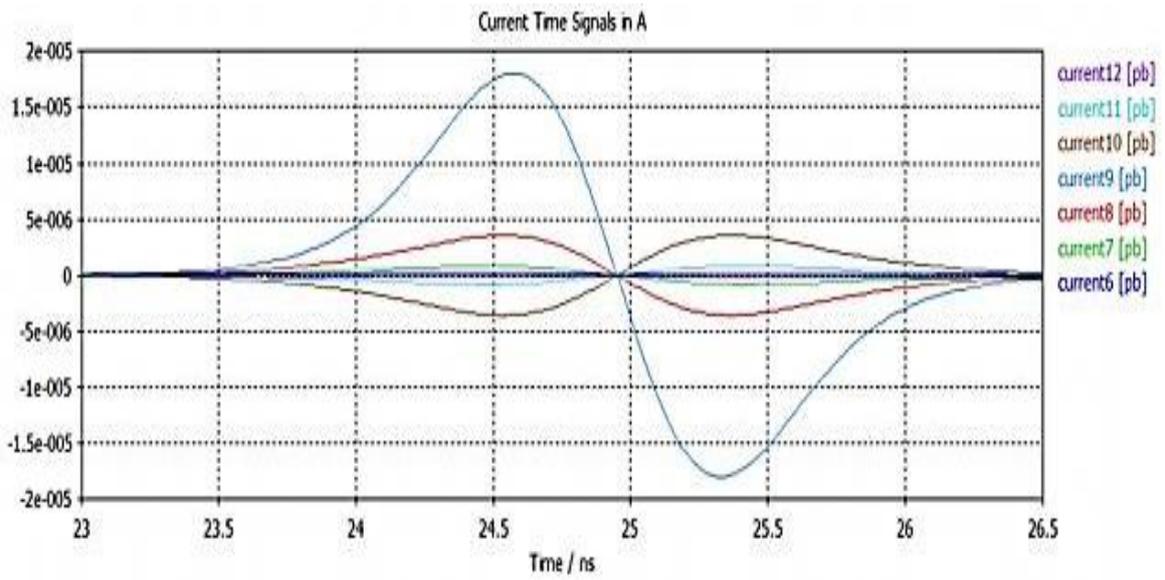

Induced current in single-loop coil with different distances

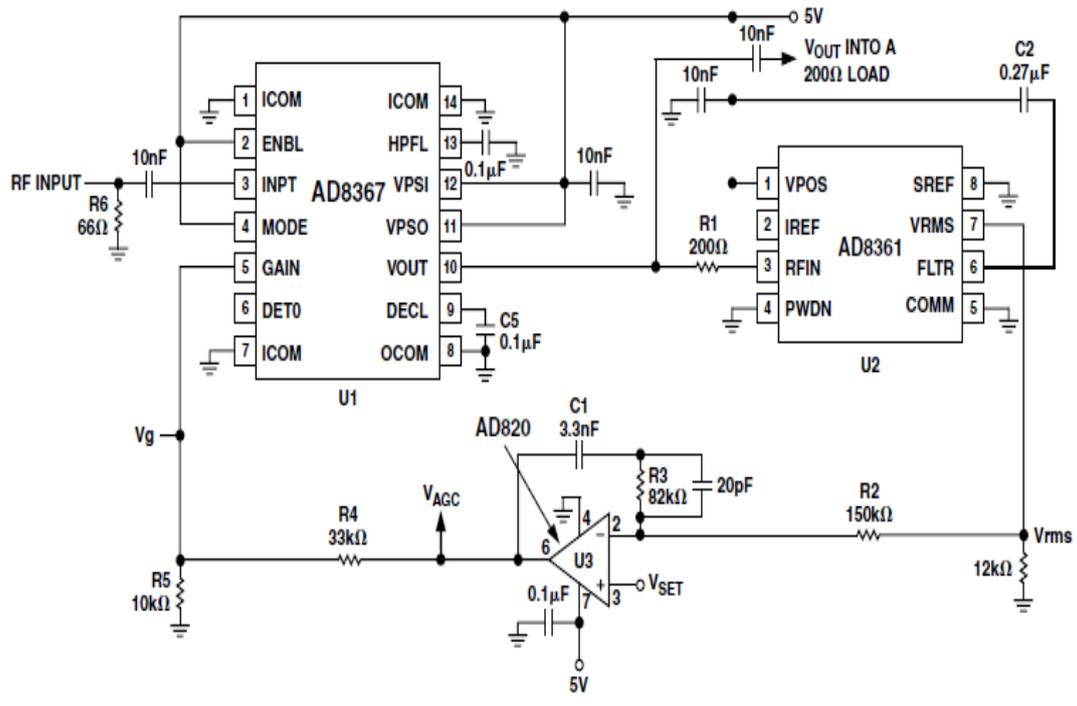

**Figure 12**

Automatic gain control for general stable RF signal amplitude

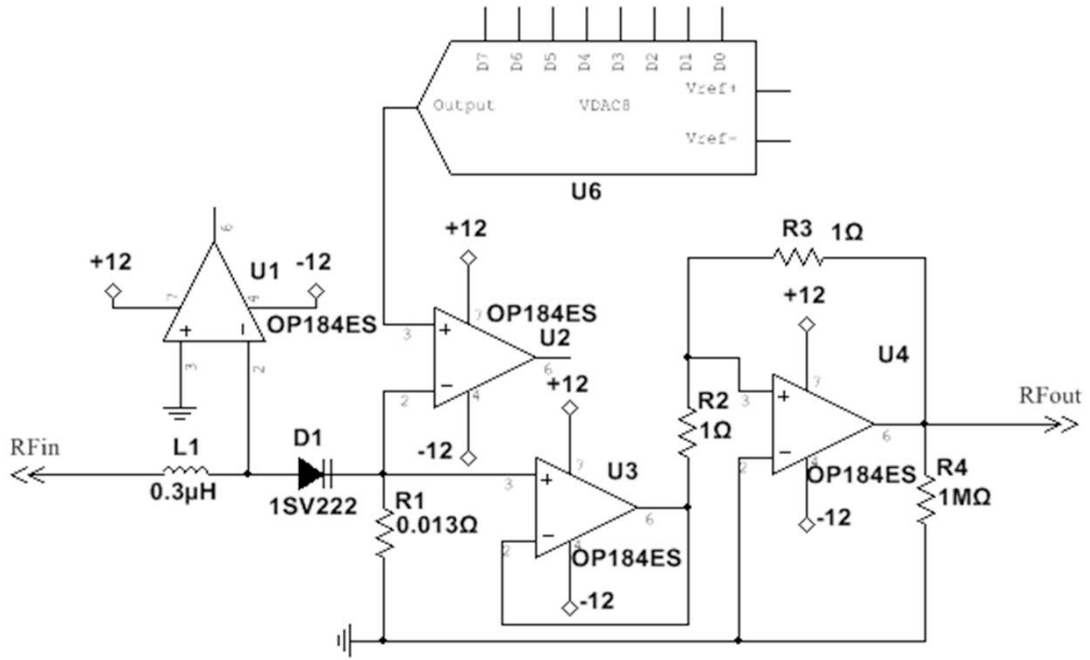

**Figure 13**

Tunable band pass filter

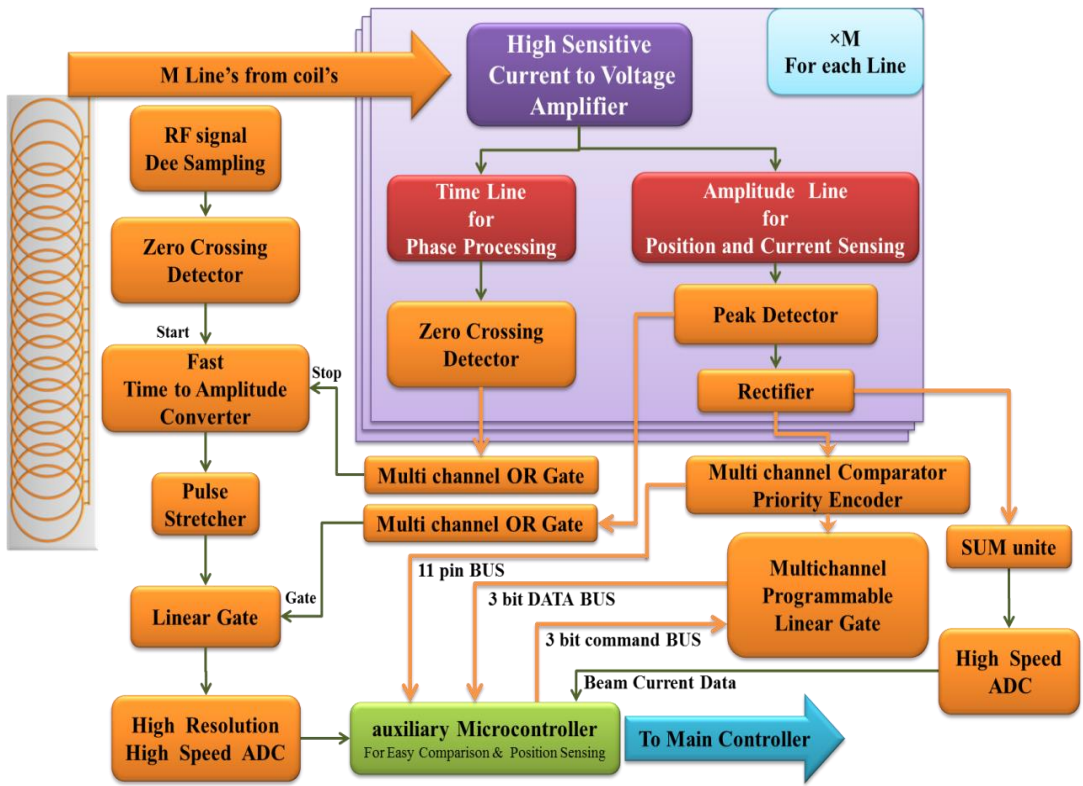

**Figure 14**

Block diagram of beam monitoring device

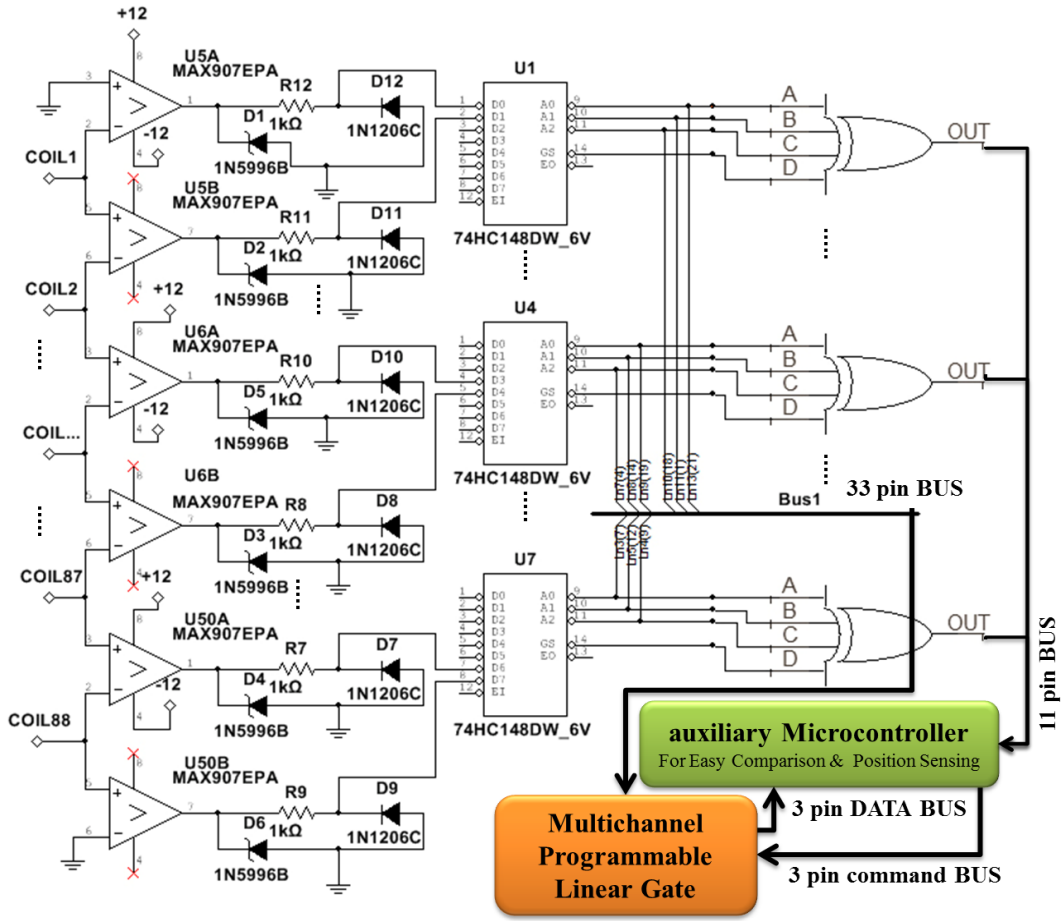

**Figure 15**

The 88 wire Parallel BUS converter to a 17 wires BUS by using Multichannel Priority Encoder Comparator and Multichannel Programmable Linear Gate